\documentclass[prd,twocolumn,showpacs,floatfix,amsmath,amssymb,floatfix]{revtex4}
\usepackage{graphicx,color,dcolumn,booktabs,bm}
\usepackage{longtable,lscape}
\usepackage{txfonts}
\usepackage{overpic}
\usepackage{amssymb}
\usepackage{indentfirst}
\usepackage{feynmf}   
\usepackage{slashed}  
\usepackage{cases}
\usepackage{color}
\usepackage{multirow}
\usepackage{epstopdf}
\usepackage{graphicx,color,dcolumn,booktabs,bm}
\usepackage{float}

\graphicspath{{Figures/}} %
\usepackage[colorlinks,
                      citecolor=blue,
                      anchorcolor=red,
                      menucolor=red,
                      linkcolor=red,
                      filecolor=red,
                      runcolor=red,
                      urlcolor=blue,
                      frenchlinks=red]{hyperref}
\begin{document}

\title{Production of $P^\Lambda_{\psi s}(4338)$ from $\Xi_b$ decay}
\author{Qi Wu$^{1}$}\email{wuqi@htu.edu.cn}
\author{Dian-Yong Chen$^{2,3}$ \footnote{Corresponding author}}\email{chendy@seu.edu.cn}
\affiliation{$^1$Institute of Particle and Nuclear Physics, Henan Normal University, Xinxiang 453007, China\\
$^2$School of Physics, Southeast University, Nanjing 210094, China\\
$^3$Lanzhou Center for Theoretical Physics, Lanzhou University, Lanzhou 730000, China}

\begin{abstract}
In the present work, we investigate the production of the newly observed $P^\Lambda_{\psi s}(4338)$ state in $\Xi_b^-$ decay, where the $P^\Lambda_{\psi s}(4338)$ is assigned as a $\Xi_c \bar{D}$ molecular state. By using an effective Lagrangian approach, we evaluate the branching fractions of $\Xi_b^-\rightarrow P^\Lambda_{\psi s}(4338)K^-$ via the triangle loop mechanism. The branching fractions of $\Xi_b^-\rightarrow P^\Lambda_{\psi s}(4338)K^-$ are in the order of $10^{-4}$; the result is compared with our previous work of $\Xi_b^-\rightarrow P^\Lambda_{\psi s}(4459)K^-$. We also predict the ratio of $P^\Lambda_{\psi s}(4459)$ and $P^\Lambda_{\psi s}(4338)$  productions in the decay $\Xi_b^- \to P^\Lambda_{\psi s} K^- \to J/\psi \Lambda K^-$. The predicted branching fractions and their ratios could be tested experimentally, which may be helpful for understanding the molecular picture of $P^\Lambda_{\psi s}(4338)$ as well as other hidden-charm pentaquark states with strangeness. Moreover, the experimental potential of observing $P^\Lambda_{\psi s}(4338)$ in the $\Xi^-_b\to K^- J/\psi \Lambda$ is discussed.
\end{abstract}

\date{\today}
\maketitle

\section{Introduction}
\label{sec:introduction}

In the past two decades, significant progress in the investigations of multiquark states has been archived on both experimental and theoretical sides (see Refs.~\cite{Chen:2016qju, Hosaka:2016pey, Lebed:2016hpi, Esposito:2016noz, Guo:2017jvc, Ali:2017jda,Olsen:2017bmm,Karliner:2017qhf,Yuan:2018inv,Dong:2017gaw,Liu:2019zoy, Chen:2022asf,Meng:2022ozq} for recent reviews). At the birth of the quark model, the notion of tetraquark and pentaquark states has been proposed in addition to the conventional mesons and baryons in 1964~\cite{Gell-Mann:1964ewy,Zweig:1964ruk}. Theoretical investigations on the possible pentaquarks composed of light quarks~\cite{Hogaasen:1978jw, Strottman:1979qu} and containing one charm quark~\cite{Lipkin:1987sk,Gignoux:1987cn} were performed after the predictions made by the quark model. The so-called $\Theta$ pentaquark, composed of $uudd\bar{s}$, was initially reported by the LEPS Collaboration in 2003~\cite{LEPS:2003wug}. However, the existence of this state has not been confirmed by the subsequent experimental measurements~\cite{Hicks:2012zz}.

In 2015, the LHCb Collaboration observed two pentaquark candidates $P^N_\psi(4380)$ and $P^N_\psi(4450)$ in the $J/\psi p$ invariant mass distribution of the decay $\Lambda_b \to K J/\psi p$~\cite{LHCb:2015yax} (henceforth, the new naming convention proposed by the LHCb Collaboration will be employed~\cite{Gershon:2022xnn}), which turns a new chapter of searching pentaquark states. Note that the hidden charm pentaquark above 4 GeV was predicted in Ref.~\cite{Wu:2010jy}. Subsequently, with the data collected in Run I and Run II, the LHCb Collaboration updated their analysis of the $J/\psi p$ invariant mass distribution of the decay $\Lambda_b \to K J/\psi p$, a new pentaquark state $P^N_\psi(4312)$ was identified, while the $P^N_\psi(4450)$ split into two structures, which were $P^N_\psi(4440)$ and $P^N_\psi(4457)$, respectively~\cite{LHCb:2019kea}. Furthermore, in 2021, the LHCb Collaboration found evidence for an additional structure $P^N_\psi(4337)$ in the $J/\psi p$ and $J/\psi \bar{p}$ systems of the decay $B^0_s\to J/\psi p\bar{p}$ with a significance in the range of 3.1 to 3.7 $\sigma$~\cite{LHCb:2021chn}.

The pentaquark candidates, $P^N_\psi(4312)$, $P^N_\psi(4337)$, $P^N_\psi(4380)$, $P^N_\psi(4440)$, and $P^N_\psi(4457)$, are all observed in the $J/\psi p$ invariant mass spectrum. Thus, their quark components are most likely to be $c\bar{c}qqq (q = u/d)$. The discovery of these states has prompted the proposal of compact pentaquark interpretations with different quark configuration~\cite{Chen:2015moa,Wang:2015epa,Wang:2019got,Ortega:2016syt,Park:2017jbn,Weng:2019ynv,Zhu:2019iwm,Deng:2022vkv}. A notable feature of these $P^N_\psi$ states is that they are slightly below or above the thresholds of the meson-baryon pairs, such as, $\Sigma_c^{(\ast)} \bar{D}^{(\ast)}$, $\Lambda_c \bar{D}^\ast$, $\chi_{c1} p$, $\psi(2S) p$. This proximity naturally leads to various molecular state explanations due to the abundant thresholds nearby~\cite{Chen:2015loa,He:2015cea,Chen:2019bip,Liu:2019tjn,Azizi:2016dhy,Huang:2015uda,Chen:2019asm,He:2019ify,He:2019rva,Zhang:2019xtu,Wang:2015qlf,
Lu:2016nnt,Shen:2016tzq,Lin:2017mtz,Guo:2019fdo,Xiao:2019mvs,Wang:2019hyc,Xu:2019zme,Lin:2019qiv,Dong:2020nwk,Wang:2019krd,Wu:2019rog,Wang:2019dsi,Yan:2021nio}. It is noteworthy that alternative explanations for $P^N_\psi(4337)$, such as cusp effect~\cite{Nakamura:2021dix} and reflection effect~\cite{Wang:2021crr}, have also been proposed.

Along the line of $P^N_\psi$ series, a pertinent question arises: does the hidden-charm pentaquark state with strangeness, denoted as $P^\Lambda_{\psi s}$, exist? The spectrum of $P^\Lambda_{\psi s}$ states has been predicted in Refs.~\cite{Hofmann:2005sw,Wu:2010vk,Anisovich:2015zqa,Wang:2015wsa,Wang:2019nvm,Park:2018oib}. Utilizing the chiral effective field theory, the authors of Ref.~\cite{Wang:2019nvm} predicted the masses of $[\Xi_c \bar{D}]_{1/2}$, $[\Xi_c \bar{D}^*]_{1/2}$, and $[\Xi_c \bar{D}^*]_{3/2}$ molecular states to be $4319.4^{+2.8}_{-3.0}$, $4456.9^{+3.2}_{-3.3}$, and $4463.0^{+2.8}_{-3.0}$ MeV, respectively. The search for $P^\Lambda_{\psi s}$ states has been suggested in the $J/\psi\Lambda$ invariant mass spectrum of the decays $\Lambda_b\rightarrow J/\psi\eta\Lambda$ \cite{Feijoo:2015kts}, $\Lambda_b\rightarrow J/\psi K^0\Lambda$ \cite{Lu:2016roh}, and $\Xi^-_b\rightarrow K^- J/\psi\Lambda$ \cite{Chen:2015sxa,Shen:2020gpw}.

Following the observations of the $P^N_\psi$ states, the LHCb Collaboration further made progress on the $P^\Lambda_{\psi s}$ states. In 2020, the LHCb Collaboration reported evidence of a pentaquark candidate $P^\Lambda_{\psi s}(4459)$ in the $J/\psi\Lambda$ invariant mass spectrum with a significance of $3.1\sigma$ in the decay $\Xi^-_b\rightarrow K^- J/\psi\Lambda$~\cite{LHCb:2020jpq}. The measured mass and width are
\begin{eqnarray}
	m =4458.8\pm2.9^{+4.7}_{-1.1} \ \mathrm{MeV}\nonumber\\
	\Gamma =17.3\pm6.5^{+8.0}_{-.57} \ \mathrm{MeV},
\end{eqnarray}
respectively. However, the $J^P$ quantum numbers of $P^N_\psi(4459)$ were not determined.

More recently, the LHCb Collaboration announced the observation of the $P^\Lambda_{\psi s}(4338)$ state in the $J/\psi\Lambda$ invariant mass spectrum of the decay $B^- \to J/\psi \Lambda\bar{p}$ with the significance more than $10\sigma$~\cite{Collaboration:2022boa}. The measured mass and width are
\begin{eqnarray}
m_{P_{cs}} &=&	4338.2\pm0.7\pm0.4 \ \mathrm{MeV}, \\
\Gamma_{P_{cs}} &=&	7.0\pm1.2\pm1.3 \ \mathrm{MeV},
\end{eqnarray}
respectively. Meanwhile, the $J^P$ quantum numbers were determined to be $J^P=\frac{1}{2}^-$.

The observations of $P^\Lambda_{\psi s}$ states have paved the way for deeper understanding the multiquark states and generated a heated discussion on their nature. Because of their proximity to $\Xi_c \bar{D}$ and $\Xi_c \bar{D}^*$ thresholds, molecular interpretations to both $P^\Lambda_{\psi s}(4338)$~\cite{Chen:2020uif, Chen:2020opr, Chen:2020kco,Liu:2020hcv,Feijoo:2022rxf} and $P^\Lambda_{\psi s}(4459)$~\cite{Karliner:2022erb,Wang:2022mxy,Yan:2022wuz,Ozdem:2022kei,Wang:2022tib,Peng:2020hql,Xiao:2021rgp,Zhu:2021lhd} have been proposed. It should be noted that the LHCb Collaboration has stated that the $P^\Lambda_{\psi s}(4459)$ structure can also be well described by a two peak structure~\cite{LHCb:2020jpq}. Inspired by this fact, the authors in Refs.~\cite{Karliner:2022erb, Wang:2022mxy} investigated the possibility that the two substructures of $P^\Lambda_{\psi s}(4459)$ corresponding to $\Xi_c \bar{D}^*$ molecular states with $J^P=\frac{1}{2}^-$ and $J^P=\frac{3}{2}^-$, which is analogous to the $P^N_\psi(4450)$ structure, contains two substructures, $P^N_\psi(4440)$ and $P^N_\psi(4457)$~\cite{LHCb:2019kea}. Additionally,  $P^\Lambda_{\psi s}(4338)$ has been interpreted as the strange partner of $P^N_\psi(4312)$. With the aim at decoding the inner structure of $P^\Lambda_{\psi s}$ states, the magnetic moment of $P^\Lambda_{\psi s}(4338)$ and $P^\Lambda_{\psi s}(4459)$ states was estimated by using the light-cone sum rules method~\cite{Ozdem:2022kei} and the constituent quark model~\cite{Wang:2022tib}. In Ref.~\cite{Burns:2022uha}, the $P^\Lambda_{\psi s}(4338)$ structure was interpreted as the triangle singularity. However, an analysis of the LHCb data on the decay $B^- \to J/\psi \Lambda\bar{p}$ reveals a pole corresponding to $P^\Lambda_{\psi s}(4338)$ at $(4339.2 \pm 1.6)-(0.9 \pm 0.4)i$ MeV~\cite{Nakamura:2022jpd}, indicating that the data does not support the $P^\Lambda_{\psi s}(4338)$ structure resulting from a kinematical effect. In Ref.~\cite{Meng:2022wgl}, the authors found that $P^\Lambda_{\psi s}(4338)$ arise from the pole well above the $\Xi^+_c D^-$ threshold or the pole well below the $\Xi^0_c \bar{D}^0$ threshold. As indicated in Refs.~\cite{Wang:2022mxy,Burns:2022uha}, regarding the $P^\Lambda_{\psi s}(4338)$ as a partner of $P^\Lambda_{\psi s}(4459)$ may be problematic, since the mass gap of the $P^\Lambda_{\psi s}$ states is different from that of the $P^N_\psi$ states.

It should be noted that the investigations on the production mechanism, as well as the mass spectrum, and decay behaviors, of the pentaquark candidates can also provide beneficial information of their internal structures. For instance, in our prior work~\cite{Wu:2019rog}, we examined  the production of $P^N_\psi(4312/4440/4457)$ in the decay $\Lambda_b\rightarrow P^N_\psi K$; the results were in good agreement with the LHCb data~\cite{LHCb:2019kea}, which supported the molecular interpretations of $P_\psi^N$ states. The production of $P^\Lambda_{\psi s}(4459)$ in $\Xi_b^-\rightarrow P^\Lambda_{\psi s} K^-$ was also estimated and the production ratio was predicted to be of the order of $10^{-4}$~\cite{Wu:2021caw}. Building on these findings, we can further explore the production mechanism of the process $\Xi_b^-\rightarrow P^\Lambda_{\psi s}(4338) K^-$ with $P^\Lambda_{\psi s}(4338)$ being a $\Xi_c \bar{D}$ molecule state by using an effective Lagrangian approach. Additionally, it has been observed that the $P^\Lambda_{\psi s}(4338)$ lies at the boundary of the phase space of the decay $B^- \to J/\psi \Lambda \bar{p}$, and the theoretical description of the resonance near the phase space boundary remains an unresolved issue. Thus, searching $P^\Lambda_{\psi s}(4338)$ in other production processes, such as $\Xi_b^-\rightarrow P^\Lambda_{\psi s}(4338) K^-$, becomes interesting and crucial.

This work is organized as follows. After the Introduction, the formulas of the $P^\Lambda_{\psi s}(4338)$ production in the decay of $\Xi_b^-$, including the effective Lagrangians and production amplitudes, are shown. The numerical results and the relevant discussions are presented in Sec.~\ref{Sec:Num}, and the last section is devoted to a short summary.

\section{Theoretical framework}
\label{sec:Sec2}

\begin{figure}[htb]
\begin{tabular}{cc}
  \centering
  \includegraphics[width=0.2\textwidth, trim=7cm 12cm 6.5cm 12cm, clip]{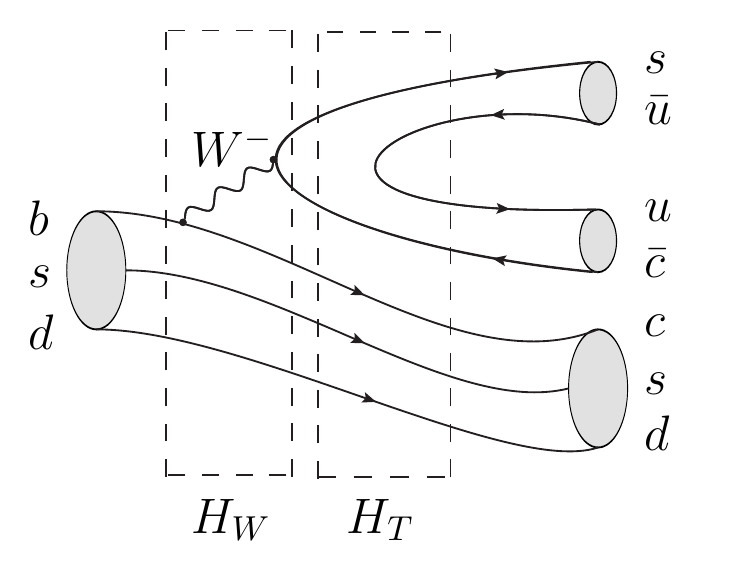}&
 \includegraphics[width=4cm]{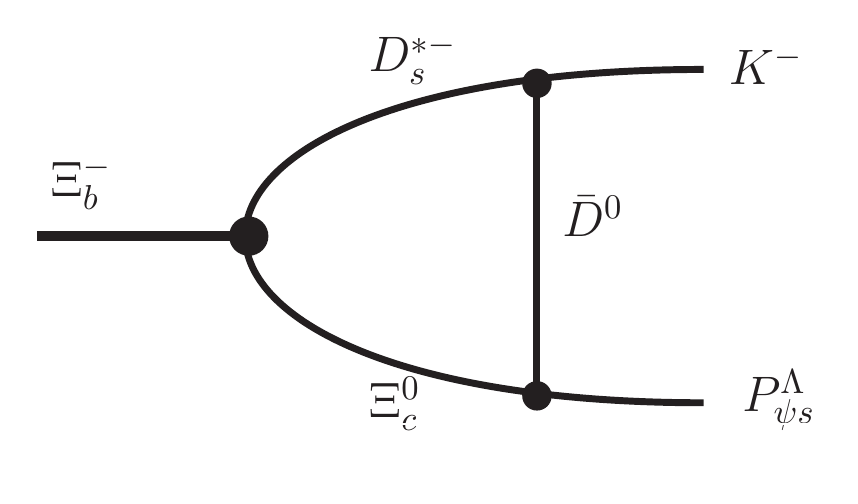}\\
 \\
 $(a)$ & $(b)$ \\
 \end{tabular}
  \caption{Diagrams contributing to $\Xi_b^-\rightarrow P^\Lambda_{\psi s}(4338) K^-$ at the quark level (a) and the hadron level (b).}\label{Fig:Tri}
\end{figure}

We can first analyze the process $\Xi_b^-\rightarrow P^\Lambda_{\psi s}(4338) K^-$ at the quark level, which is illustrated in Fig.~\ref{Fig:Tri}(a). From the phenomenological perspective, this decay occurs in two steps, which are
\begin{itemize}
	\item  the bottom quark transits to charm quark by emitting a $W^-$ boson that couples to the $\bar{c} s$ quark pair consequently;
	\item the $\bar{c} s$ quark pair and the $u\bar{u}$  created from vacuum transit into $K^-$ and $\bar{D}$, and then $\bar{D}$ and $\Xi_c$ form a bound state, i.e., $P^\Lambda_{\psi s}(4338)$.
\end{itemize}
Here we phenomenologically describe the above two steps by the operators $\mathcal{H}_W$ and $\mathcal{H}_T$, respectively. Then, the amplitude of the decay process $\Xi_b^- \to P_{\psi s}^\Lambda(4338) K^-$ can be expressed,
\begin{eqnarray}
	\langle P_{\psi s}^\Lambda(4338) K^- \left|\mathcal{H}_T \mathcal{H}_W \right|\Xi_b^- \rangle.
\end{eqnarray}
The estimation of the above amplitude at the quark level is rather difficult, and in the present work, we try to evaluate this amplitude at the hadron level by inserting a complete basis formed by a baryon and a meson between $\mathcal{H}_T$ and $\mathcal{H}_W$, which is
 \begin{eqnarray}
	&&\langle P_{\psi s}^\Lambda(4338) K^- \left|\mathcal{H}_T \mathcal{H}_W \right|\Xi_b^- \rangle\nonumber\\
	&& \quad \qquad=\sum_{B,M}
	\langle P_{\psi s}^\Lambda(4338) K^- |\mathcal{H}_T  | BM\rangle \langle BM | \mathcal{H}_W |\Xi_b^- \rangle.
\end{eqnarray}
In principle, all possible bases that can connect the initial $\Xi_b^-$ and final $P_{\psi s}^\Lambda(4338) K^-$ should be taken into account. In the present work, the $P_{\psi s}^\Lambda(4338)$ is regarded as a $\Xi_c^0 \bar{D}^0$ molecule, indicating  the dominance of the coupling between  $P_{\psi s}^\Lambda(4338)$ and its components $\Xi_c^0 \bar{D}^0$. Therefore, at the hadron level, the contributions from the diagram in Fig.~\ref{Fig:Tri}(b), where the initial $\Xi_b^-$ and final $P_{\psi s}^\Lambda(4338) K^-$ are connected by $\Xi_c^0 D_s^{\ast -}$ by exchanging a $\bar{D}^0$ meson, are expected to be dominant.

In the present work, we employ an effective Lagrangian approach to estimate the diagram illustrated in Fig.~\ref{Fig:Tri}(b). The Lagrangian for the weak vertex $\Xi_b \Xi_c D^{\ast}_s$ reads \cite{Cheng:1996cs}
\begin{eqnarray}
\mathcal{L}_{\Xi_b \Xi_c D^\ast_s} = D^{\ast\mu}_s \bar{\Xi}_c (A_1 \gamma_\mu \gamma_5+A_2 p_{2\mu}\gamma_5+B_1 \gamma_\mu+B_2 p_{2\mu})\Xi_b,\ \
\label{eq:1}
\end{eqnarray}
where $A_1,\ A_2,\ B_1$, and $B_2$ are the recombinations of the form factors $g_{1,2}$ and $f_{1,2}$, which are
\begin{eqnarray}
  A_1 &=& -\lambda a_1 f_{D^\ast_s}m_1[g_1+g_2(m-m_2)],\nonumber\\
  A_2&=& -2\lambda a_1 f_{D^\ast_s}m_1 g_2,\nonumber\\
  B_1 &=& \lambda a_1 f_{D^\ast_s}m_1[f_1-f_2(m+m_2)],\nonumber\\
   B_2&=& 2\lambda a_1 f_{D^\ast_s}m_1 f_2,
\end{eqnarray}
where $\lambda=\frac{G_F}{\sqrt{2}}V_{cb}V_{cs}$ and $a_1=1.07$~\cite{Li:2012cfa}. $m$, $m_1$, and $m_2$ are the masses of $\Xi_b^-$, $D^{\ast-}_s$, and $\Xi_c^0$ respectively. $f_{D^*_s}=0.247$ is the decay constant of the $D_{s}^\ast$, which is estimated by twisted-mass lattice QCD~\cite{Carrasco:2014poa}.

The transition form factors of $\Xi_b^-\to \Xi_c^0$ could be parametrized in the form \cite{Cheng:1996cs}
\begin{eqnarray}
f_i(Q^2)=\frac{f_i(0)}{(1-Q^2/m^2_V)^2},\ \ \ g_i(Q^2)=\frac{g_i(0)}{(1-Q^2/m^2_A)^2},
\label{Eq:A1}
\end{eqnarray}
where $m_V (m_A)$ is the pole mass of the vector (axial-vector) meson. In Table~\ref{Tab:FFs}, we collect the parameters related to the transition form factors of $\Xi_b^-\to \Xi_c^0$ \cite{Cheng:1996cs}.

\begin{table}[htb]
\begin{center}
\caption{The values of the parameters $f_i(0)$ and $g_i(0)$ in the form factors of $\Xi_b^-\to \Xi_c^0$ transition\cite{Cheng:1996cs}.}\label{Tab:FFs}
  \setlength{\tabcolsep}{2.4mm}{
\begin{tabular}{ccccccc}
  \toprule[1pt]
 ~~~ Parameter~~~ & ~~~ Value ~~~& ~~~Parameter~~~ & ~~~Value~~~\\
  \midrule[1pt]
 $f_1(0)$ &   0.533  & $g_1(0)$ & 0.580 \\
 $m_{\Xi_b} f_2(0)$  & -0.124  & $m_{\Xi_b}g_2(0)$    & -0.019   \\
 $m_V$  & $6.34$ GeV  & $m_A$ & $6.73$ GeV \\
  \bottomrule[1pt]
\end{tabular}}
\end{center}
\end{table}

Based on the SU(4) symmetry, the effective Lagrangian of $D^{\ast}_s D K$ is~\cite{Azevedo:2003qh}
\begin{eqnarray}
\mathcal{L}_{KDD^{\ast}_s}&=&i g_{KDD^{\ast}_s}D^{\ast\mu}_s[\bar{D}\partial_\mu\bar{K}-(\partial_\mu\bar{D})\bar{K}]+\mathrm{H.c.},
\end{eqnarray}
with $g_{KDD^{\ast}_s}=5.0$~\cite{Azevedo:2003qh}. In the molecular scenario, $P^\Lambda_{\psi s}(4338)$ is considered to be a molecular state composed of $\bar{D} \Xi_c$ with $J^P =1/2^-$ and the $S$-wave coupling of $P^\Lambda_{\psi s}(4338)$ and $\Xi_c \bar{D}$ is~\cite{Zou:2002yy}
\begin{eqnarray}
\mathcal{L}_{P_{\psi s}^\Lambda\Xi_c \bar{D}} &=& g_{P_{\psi s}^\Lambda \Xi_c D}\bar{\Xi}_c P_{\psi s} \bar{D}^0 +\mathrm{H.c.}, \nonumber\\
\end{eqnarray}
where $g_{P_{\psi s}^\Lambda\Xi_c D}$ is the coupling constant of $P_{\psi s}^\Lambda $ and $\Xi_c D$. As an $S$-wave shallow bound state, the coupling constants of $P_{\psi s}^\Lambda(4338)$ and its components $ \Xi_c \bar{D}$ could be estimated under nonrelativistic conditions~\cite{Weinberg:1965zz,Baru:2003qq},
\begin{eqnarray}
	g_{P_{\psi s}^\Lambda\Xi_c D}^2 =\frac{4\pi}{4m_0 m_2} \frac{(m_1+m_2)^{5/2}}{(m_1m_2)^{1/2}} \sqrt{32 E_b},
	\label{Eq:CP-Non}
\end{eqnarray}
where $m_0$, $m_1$, and $m_2$ are the masses of $P^\Lambda_{\psi s}(4338)$, $D$, and $\Xi_c$, respectively, and $E_b=m_1+m_2-m_0$ is the binding energy of the $S$-wave shallow bound state. With the measured mass of $P^\Lambda_{\psi s}(4338)$, the coupling constant is estimated to be $g=1.21$.

With the effective Lagrangians listed above, we can obtain the amplitudes of $\Xi_b^-(p)\rightarrow D^{\ast-}_s(p_1)\Xi_c^0(p_2)[D(q)]\rightarrow K^-(p_3)P^\Lambda_{\psi s}(4338)(p_4)$ corresponding to Fig.~\ref{Fig:Tri}(b), which is
\begin{eqnarray}
\mathcal{M}&=&i^3 \int\frac{d^4 q}{(2\pi)^4}\Big [g_{p_{\psi s}\Xi_c D}\bar{u}(p_4)\Big]\Big(p_2\!\!\!\!\!\slash+m_2\Big) \nonumber\\
&&\times  \Big[ \Big(A_1 \gamma_\alpha \gamma_5+A_2 p_{2\alpha}\gamma_5+B_1 \gamma_\alpha+B_2 p_{2\alpha} \Big) u(p) \Big]\nonumber\\
&&\times  \Big[ig_{KD D^\ast_s}i(p_{3\beta}-q_\beta)\Big] \frac{-g^{\alpha\beta}+p_1^{\alpha}p_1^\beta/m^2_1}{p^2_1-m^2_1}\nonumber\\
&&\times \frac{1}{p^2_2-m^2_2}\frac{1}{q^2-m^2_E}\mathcal{F}(q^2,m_{E}^2).
\end{eqnarray}

In the present work, a form factor in the monopole form is introduced to depict the internal structure of the exchanged charmed meson and avoiding the divergence in the loop integration. The concrete expression of the form factor is
\begin{eqnarray}
\mathcal{F}(q^2,m^2) =\frac{m^2 -\Lambda^2}{q^2-\Lambda^2},\label{Eq:FFs}
\end{eqnarray}
where $\Lambda=m+\alpha \Lambda_{\mathrm{QCD}}$~\cite{Cheng:2004ru} with $\Lambda_{\mathrm{QCD}}=220$ MeV. $\alpha$ is a model parameter, which should be of the order of unity~\cite{Tornqvist:1993vu,Locher:1993cc,Li:1996yn}.

\section{Numerical Results and Discussions}
\label{Sec:Num}

\begin{figure}[htb]
\centering
\includegraphics[width=0.65\textwidth, trim=2cm 17cm 4cm 4cm, clip]{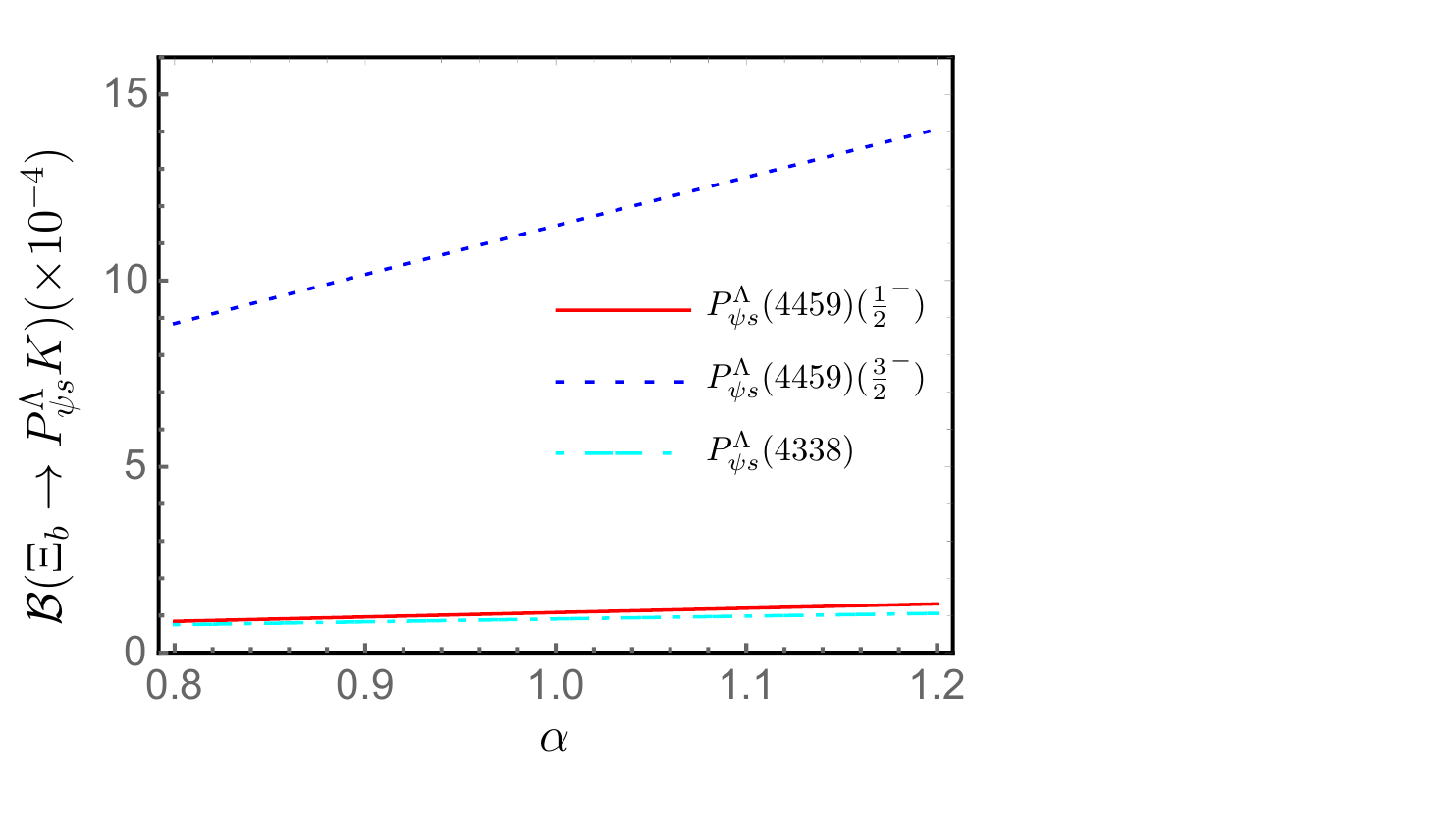}
\caption{The branching fractions of $\Xi_b^-\rightarrow P^\Lambda_{\psi s}(4338)K^-$ depending on the parameter $\alpha$. For comparison, we also present the branching fraction of $\Xi_b^-\rightarrow P^\Lambda_{\psi s}(4459)K^-$ with different $J^P$ assignments for $P^\Lambda_{\psi s}(4459)$~\cite{Wu:2019rog}.}
\label{fig:BR}
\end{figure}

With the above preparations, we can estimate the branching fractions of $\Xi_b^- \to P^\Lambda_{\psi s}(4338)K^-$. The $\alpha$ dependence of the branching fraction is shown in Fig.~\ref{fig:BR}, where we also present our estimations of the branching fractions for $\Xi_b^- \to P^\Lambda_{\psi s}(4459)K^-$ with different $J^P$ assignments for $P^\Lambda_{\psi s}(4459)$~\cite{Wu:2019rog}. Here, we vary the model parameter $\alpha$ from 0.8  to 1.2, which is the same as that of our previous investigations of $P^N_\psi(4312/4440/4457)$~\cite{Wu:2019rog} and $P^\Lambda_{\psi s}(4459)$~\cite{Wu:2021caw} productions. From the figure, one can find the estimated branching ratios increase with the increasing of model parameter $\alpha$. In particular, in the considered $\alpha$ range, the branching ratio of $\Xi_b^- \to P^\Lambda_{\psi s}(4338)K^-$ is estimated to be $(0.75\sim 1.06)\times 10^{-4}$. By comparing the branching ratios of $\Xi_b^- \to P^\Lambda_{\psi s}(4338)K^- $ and $\Xi_b^- \to P^\Lambda_{\psi s}(4459)K^-$, we find that the branching ratio of $\Xi_b^-\rightarrow P^\Lambda_{\psi s}(4338)K^-$ is almost the same as that of $\Xi_b^-\rightarrow P^\Lambda_{\psi s}(4459)(\frac{1}{2}^-)K^-$, while it is about 1 order of magnitude smaller than that of $\Xi_b^-\rightarrow P^\Lambda_{\psi s}(4459)(\frac{3}{2}^-)K^-$.

\begin{figure}[htb]
\centering
\includegraphics[width=0.68\textwidth, trim=2cm 17.5cm 4.5cm 4.25cm, clip]{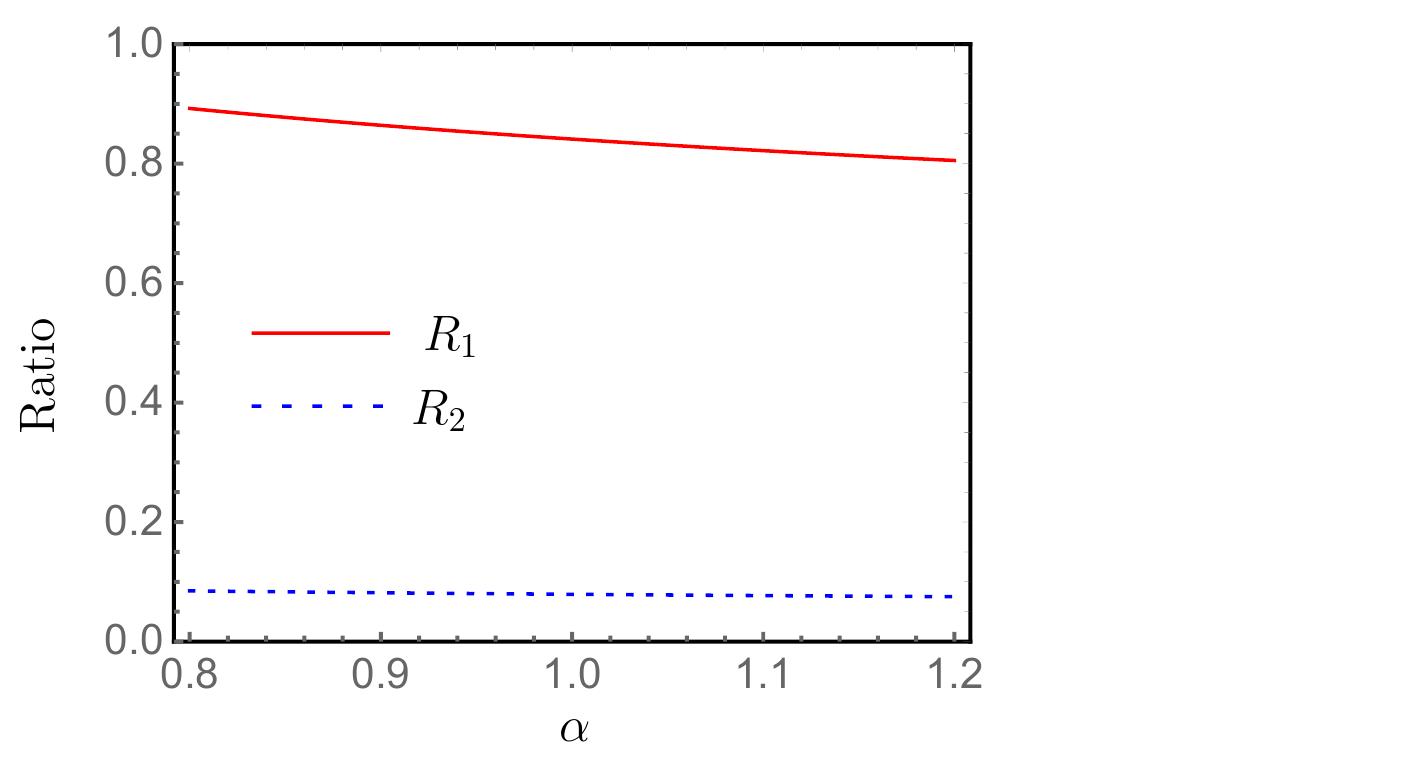}
\caption{The branching fraction ratios depending on the model parameter $\alpha$.}
\label{fig:ratio}
\end{figure}

To further check the relations of the production rates of $P^\Lambda_{\psi s}(4338)$ and $P^\Lambda_{\psi s}(4459)$ states in $\Xi_b^-$ decays, we define the following ratios:
\begin{eqnarray}
R_1=\frac{\mathcal{B}[\Xi_b^-\rightarrow P^\Lambda_{\psi s}(4338)K^-]}{{\mathcal{B}[\Xi_b^-\rightarrow P^\Lambda_{\psi s}(4459)(\frac{1}{2}^-)K^-]}},\nonumber\\
R_2=\frac{\mathcal{B}[\Xi_b^-\rightarrow P^\Lambda_{\psi s}(4338)K^-]}{{\mathcal{B}[\Xi_b^-\rightarrow P^\Lambda_{\psi s}(4459)(\frac{3}{2}^-)K^-]}}.
\end{eqnarray}
The $\alpha$ dependence of these ratios is presented in Fig.~\ref{fig:ratio}. The ratio $R_1$ is estimated to be $0.81 \sim 0.89$, while the ratio $R_2$ is evaluated to be $(7.5 \sim 8.5)\times 10^{-2}$. Our estimations find that these two ratios are almost  independent on the model parameter, which can be  tested by further experimental measurements from the LHCb Collaboration.

\begin{figure}[t]
  \centering
 \includegraphics[width=0.43\textwidth, trim=2cm 16cm 5cm 2cm, clip]{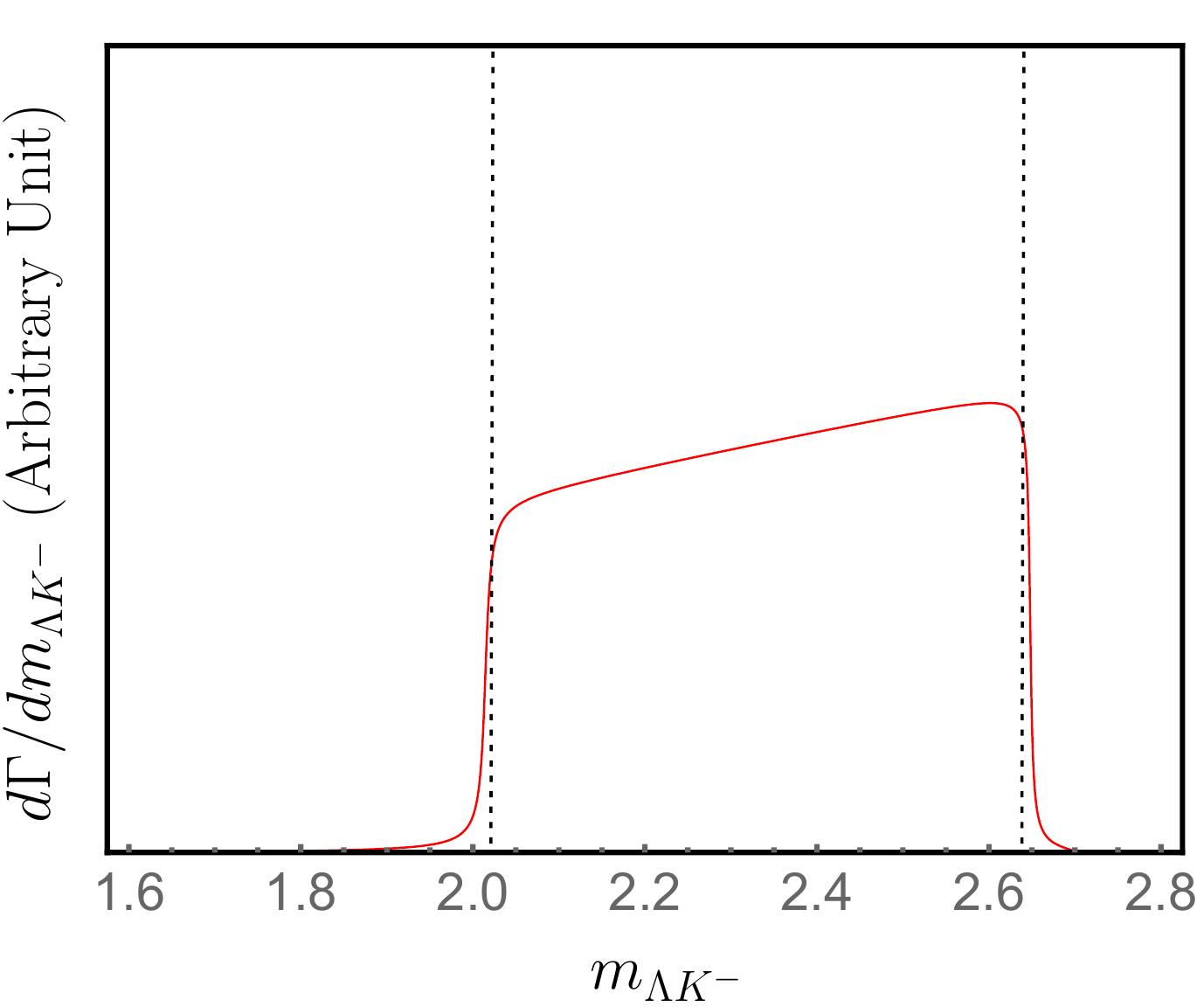}
  \caption{The signal of $P_{\psi s}^\Lambda (4338)$ in the $\Lambda K^-$ invariant mass distributions.}\label{fig:projection}
\end{figure}

\begin{figure}[t]
  \centering
 \includegraphics[width=1.00\linewidth]{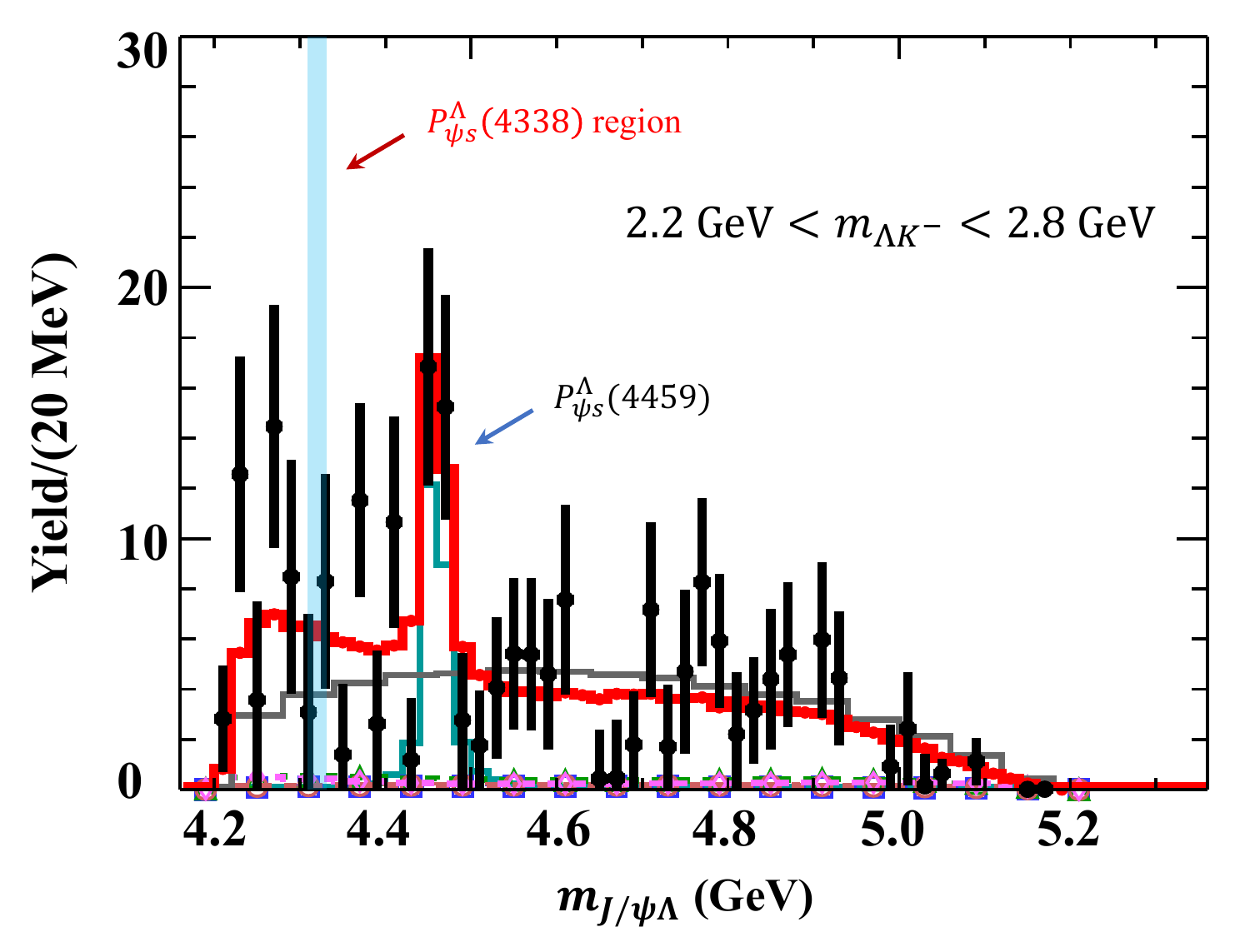}
  \caption{The $J/\psi \Lambda$ invariant mass distributions of $\Xi^-_b \to J/\psi \Lambda K^-$ with $2.2~\mathrm{GeV}<m_{\Lambda K^-}<2.8~\mathrm{GeV}$ reported by the LHCb Collaboration~\cite{LHCb:2020jpq}. The cyan band is the signal region of $P_{\psi s}^\Lambda (4338)$.}\label{fig:LHCbdata}
\end{figure}

Before the end of this work, we would like to discuss the experimental potential of observing $P^\Lambda_{\psi s}(4338)$ in the decay $\Xi^-_b\to K^- J/\psi \Lambda$. In Ref.~\cite{Wu:2010vk}, the branching ratio of $P_{\psi s}^\Lambda(4338) \rightarrow J/\psi\Lambda$ is estimated to be $(9.74^{+2.69}_{-2.36}) \times 10^{-2}$. Considering the small width of $P^\Lambda_{\psi s}(4338)$, and together with present estimation, one has $\mathcal{B}(\Xi_b^-\rightarrow P^\Lambda_{\psi s}(4338)K^-\rightarrow J/\psi\Lambda K^-)\simeq\mathcal{B}(\Xi^-_b\rightarrow P^\Lambda_{\psi s}(4338)K^-)\times\mathcal{B}(P^\Lambda_{\psi s}(4338)\rightarrow J/\psi\Lambda)=\left( 8.84^{+2.84}_{-2.63} \right) \times 10^{-6}$. In Ref.~\cite{Wu:2021caw}, with the help of the relevant experimental measurement, we obtain $\mathcal{B}(\Xi^-_b \to P^\Lambda_{\psi s}(4459) K^- \to J/\psi \Lambda K^-)=\left( 6.25^{+5.98}_{-4.98} \right) \times 10^{-6}$. Thus, the ratio of the fit fractions of $P^\Lambda_{\psi s}(4459)$ and $P^\Lambda_{\psi s}(4338)$ in $\Xi_b^- \to J/\psi \Lambda K^-$ is predicted to be
\begin{eqnarray}
\frac{\mathcal{B}(\Xi_b^- \to P^\Lambda_{\psi s}(4459) K^- \to J/\psi \Lambda K^-)}{\mathcal{B}(\Xi_b^- \to P^\Lambda_{\psi s}(4338) K^- \to J/\psi \Lambda K^-)} =0.71^{+0.71}_{-0.60}.
\end{eqnarray}
As shown in Fig.~\ref{fig:projection}, the signal of $P_{\psi s}^\Lambda (4338)$ can project into the $\Lambda K^-$ invariant mass spectroscopy mainly in the region $2.0~\mathrm{GeV}<m_{\Lambda K^-}<2.6~\mathrm{GeV}$. In Ref.~\cite{LHCb:2020jpq}, the LHCb Collaboration reported the $J/\psi \Lambda $ invariant mass distributions with $2.2~\mathrm{GeV}<m_{\Lambda K^-} < 2.8 ~\mathrm{GeV}$ of the decay $\Xi_b^- \to J/\psi \Lambda K^-$, where the pentaquark state $P_{\psi s}^\Lambda (4459)$ was observed. In Fig.~\ref{fig:LHCbdata}, we present the experimental data from the LHCb Collaboration, from which one can find the data between 4.2 and 4.4 GeV have very large uncertainties. In addition, the bin size of the experimental data is 20 MeV, which is much larger than the width of $P_{\psi s}^\Lambda (4338)$. Further precise measurements of the $J/\psi \Lambda $ invariant mass distributions with $2.0~\mathrm{GeV}<m_{\Lambda K^-} < 2.6 ~\mathrm{GeV}$ may shed light on the production of $P_{\psi s}^\Lambda (4338)$ in the decay $\Xi_b^- \to J/\psi \Lambda K^-$.

\section{Summary}
\label{Sec:Summary}

The notion of pentaquark can retrospect to the birth of the quark model. After that, searching for possible pentaquarks keeps intriguing for both experimentalists and theorist. The first pentaquark candidate is the $\Theta^+$ composed of $uudd\bar{s}$ which were first reported by the LEPS Collaboration in 2003. However, the subsequent experiments did not confirm the existence of these pentaquark states. The breakthrough was made by the LHCb Collaboration in the year of 2015, and further updated in 2019, when a series of  hidden-charm pentaquark, $P^N_\psi$ states, were observed.

After the observations of $P^N_\psi$, the LHCb Collaboration searched for the $P^\Lambda_{\psi s}$ states in the decay $\Xi_b^-\rightarrow J/\psi \Lambda K^-$ and $B^- \to J/\psi \Lambda\bar{p}$. In the former process, the first hidden-charm pentaquark state candidate with strangeness $P^\Lambda_{\psi s}(4459)$ was discovered. In the latter process, anther hidden-charm pentaquark state with strangeness, $P^\Lambda_{\psi s}(4338)$, was discovered with definitely $J^P$ quantum numbers. Theorists have proposed various models to understand their spectrum, decay behaviors, and production mechanism.

In the present work, we studied the production process $\Xi_b^-\to P^\Lambda_{\psi s}(4338)  K^-$ with an effective Lagrangian approach. We first analyze the process $\Xi_b^-\rightarrow P^\Lambda_{\psi s}K^-$ at the quark level and then estimate it at the hadron level via triangle loop mechanism. The branching ratio of $\Xi_b^-\rightarrow P^\Lambda_{\psi s}(4338)K^-$ is predicted to be of the order of $10^{-4}$. By comparison with our previous work of $\Xi_b^-\rightarrow P^\Lambda_{\psi s}(4459)K^-$, the branching ratio of $\Xi_b^-\rightarrow P^\Lambda_{\psi s}(4338)K^-$ is very close to that of $\Xi_b^-\rightarrow P^\Lambda_{\psi s}(4459)(\frac{1}{2}^-)K^-$, and of 1 order of magnitude smaller than that of $\Xi_b^-\rightarrow P^\Lambda_{\psi s}(4459)(\frac{3}{2}^-)K^-$. The ratios of $P^\Lambda_{\psi s}(4338)$ and $P^\Lambda_{\psi s}(4459)$ with $J^P=\frac{1}{2}^-$ and $J^P=\frac{3}{2}^-$ in $\Xi_b^-\rightarrow P^\Lambda_{\psi s}K^-$ are estimated, the predicted ratios are almost independent on the model parameter and can be tested by further experimental measurements from the LHCb Collaboration.

At last, we discussed the experimental potential of observing $P^\Lambda_{\psi s}(4338)$ in the decay $\Xi^-_b\to K^- J/\psi \Lambda$. The ratios of the fit fractions of $P^\Lambda_{\psi s}(4459)$ and $P^\Lambda_{\psi s}(4338)$ in $\Xi_b^- \to J/\psi \Lambda K^-$ are estimated, which is smaller than unity. We explored the scope of the signal of $P_{\psi s}^\Lambda (4338)$ projecting into the $\Lambda K^-$ invariant mass spectroscopy and compared the present estimations with the LHCb data. Then, we suggested that the further precise measurements of the $J/\psi \Lambda$ invariant mass distributions with $2.0~\mathrm{GeV}<m_{\Lambda K^-} < 2.6 ~\mathrm{GeV}$ may shed light on the production of $P_{\psi s}^\Lambda (4338)$ in the decay $\Xi_b^- \to J/\psi \Lambda K^-$.

\bigskip
\noindent
\begin{center}
	{\bf ACKNOWLEDGEMENTS}\\
\end{center}
This work is supported by the National Natural Science Foundation of China under Grants No.~12175037 and No.~12335001.
\bigskip



\begin{thebibliography}{0}

\bibitem{Chen:2016qju}
  H.-X.~Chen, W.~Chen, X.~Liu, and S.-L.~Zhu,
  Phys.\ Rep.\  {\bf 639}, 1 (2016).


\bibitem{Hosaka:2016pey}
  A.~Hosaka, T.~Iijima, K.~Miyabayashi, Y.~Sakai, and S.~Yasui,
 Prog. Theor. Exp. Phys. {\bf 2016}, 062C01 (2016).


\bibitem{Lebed:2016hpi}
  R.~F.~Lebed, R.~E.~Mitchell, and E.~S.~Swanson,
  Prog.\ Part.\ Nucl.\ Phys.\  {\bf 93}, 143 (2017).


\bibitem{Esposito:2016noz}
  A.~Esposito, A.~Pilloni, and A.~D.~Polosa,
  Phys.\ Rep.\  {\bf 668}, 1 (2016).


\bibitem{Guo:2017jvc}
F.~K.~Guo, C.~Hanhart, U.~G.~Mei\ss{}ner, Q.~Wang, Q.~Zhao, and B.~S.~Zou,
Rev. Mod. Phys. \textbf{90}, 015004 (2018); 90, 015004(E) (2022).


\bibitem{Ali:2017jda}
  A.~Ali, J.~S.~Lange, and S.~Stone,
  Prog.\ Part.\ Nucl.\ Phys.\  {\bf 97}, 123 (2017).


\bibitem{Olsen:2017bmm}
  S.~L.~Olsen, T.~Skwarnicki, and D.~Zieminska,
  Rev.\ Mod.\ Phys.\  {\bf 90}, 015003 (2018).


\bibitem{Karliner:2017qhf}
  M.~Karliner, J.~L.~Rosner, and T.~Skwarnicki,
  Annu.\ Rev.\ Nucl.\ Part.\ Sci.\  {\bf 68}, 17 (2018).


\bibitem{Yuan:2018inv}
  C.-Z.~Yuan,
  Int.\ J.\ Mod.\ Phys.\ A {\bf 33}, 1830018 (2018).

\bibitem{Dong:2017gaw}
Y.~Dong, A.~Faessler, and V.~E.~Lyubovitskij,
Prog. Part. Nucl. Phys. \textbf{94}, 282 (2017).


\bibitem{Liu:2019zoy}
Y.~R.~Liu, H.~X.~Chen, W.~Chen, X.~Liu, and S.~L.~Zhu,
Prog. Part. Nucl. Phys. \textbf{107},237 (2019).

\bibitem{Chen:2022asf}
H.~X.~Chen, W.~Chen, X.~Liu, Y.~R.~Liu, and S.~L.~Zhu,
Rep. Prog. Phys. \textbf{86}, 026201 (2023).

\bibitem{Meng:2022ozq}
L.~Meng, B.~Wang, G.~J.~Wang, and S.~L.~Zhu,
Phys. Rep.1019,1 (2023).

\bibitem{Gell-Mann:1964ewy}
M.~Gell-Mann,
Phys. Lett. \textbf{8},214 (1964).

\bibitem{Zweig:1964ruk}
G.~Zweig, An SU(3) model for strong interaction symmetry and its breaking. Version 1, CERN-TH-401.

\bibitem{Hogaasen:1978jw}
H.~Hogaasen and P.~Sorba,
Nucl. Phys. B \textbf{145}, 119 (1978).

\bibitem{Strottman:1979qu}
D.~Strottman,
Phys. Rev. D \textbf{20}, 748 (1979).

\bibitem{Lipkin:1987sk}
H.~J.~Lipkin,
Phys. Lett. B \textbf{195}, 484 (1987).

\bibitem{Gignoux:1987cn}
C.~Gignoux, B.~Silvestre-Brac and J.~M.~Richard,
Phys. Lett. B \textbf{193}, 323 (1987).

\bibitem{LEPS:2003wug}
T.~Nakano \textit{et al.} (LEPS Collaboration),
Phys. Rev. Lett. \textbf{91}, 012002 (2003).

\bibitem{Hicks:2012zz}
K.~H.~Hicks,
Eur. Phys. J. H \textbf{37}, 1 (2012).

\bibitem{LHCb:2015yax}
R.~Aaij \textit{et al.} (LHCb Collaboration),
Phys. Rev. Lett. \textbf{115}, 072001 (2015).

\bibitem{Gershon:2022xnn}
T.~Gershon (LHCb Collaboration), arXiv:2206.15233.


\bibitem{Wu:2010jy}
J.~J.~Wu, R.~Molina, E.~Oset, and B.~S.~Zou,
Phys. Rev. Lett. \textbf{105}, 232001 (2010).

\bibitem{LHCb:2019kea}
R.~Aaij \textit{et al.} (LHCb Collaboration),
Phys. Rev. Lett. \textbf{122}, 222001 (2019).

\bibitem{LHCb:2021chn}
R.~Aaij \textit{et al.} (LHCb Collaboration),
Phys. Rev. Lett. \textbf{128}, 062001 (2022).



\bibitem{Chen:2015moa}
  H.~X.~Chen, W.~Chen, X.~Liu, T.~G.~Steele, and S.~L.~Zhu,
  Phys.\ Rev.\ Lett.\  {\bf 115}, 172001 (2015).

\bibitem{Wang:2015epa}
  Z.~G.~Wang,
  Eur.\ Phys.\ J.\ C {\bf 76}, 70 (2016).

\bibitem{Wang:2019got}
  Z.~G.~Wang,
  Int.\ J.\ Mod.\ Phys.\ A {\bf 35}, 2050003 (2020).

\bibitem{Ortega:2016syt}
  P.~G.~Ortega, D.~R.~Entem, and F.~Fern\'{a}ndez,
  Phys.\ Lett.\ B {\bf 764}, 207 (2017).

\bibitem{Park:2017jbn}
  W.~Park, A.~Park, S.~Cho, and S.~H.~Lee,
  Phys.\ Rev.\ D {\bf 95}, 054027 (2017).


\bibitem{Weng:2019ynv}
  X.~Z.~Weng, X.~L.~Chen, W.~Z.~Deng, and S.~L.~Zhu,
  Phys.\ Rev.\ D {\bf 100}, 016014 (2019).


\bibitem{Zhu:2019iwm}
  R.~Zhu, X.~Liu, H.~Huang, and C.~F.~Qiao,
  Phys.\ Lett.\ B {\bf 797}, 134869 (2019).

\bibitem{Deng:2022vkv}
C.~R.~Deng,
Phys. Rev. D \textbf{105}, 116021 (2022).

\bibitem{Chen:2015loa}
  R.~Chen, X.~Liu, X.~Q.~Li, and S.~L.~Zhu,
  Phys.\ Rev.\ Lett.\  {\bf 115}, 132002 (2015).

\bibitem{He:2015cea}
  J.~He,
  Phys.\ Lett.\ B {\bf 753}, 547 (2016).

\bibitem{Liu:2019tjn}
  M.~Z.~Liu, Y.~W.~Pan, F.~Z.~Peng, M.~S\'{a}nchez S\'{a}nchez, L.~S.~Geng, A.~Hosaka, and M.~Pavon Valderrama,
  Phys.\ Rev.\ Lett.\  {\bf 122}, 242001 (2019).


\bibitem{Huang:2015uda}
  H.~Huang, C.~Deng, J.~Ping, and F.~Wang,
  Eur.\ Phys.\ J.\ C {\bf 76}, 624 (2016).


\bibitem{Chen:2019asm}
  R.~Chen, Z.~F.~Sun, X.~Liu, and S.~L.~Zhu,
  Phys.\ Rev.\ D {\bf 100}, 011502 (2019).

\bibitem{He:2019ify}
  J.~He,
  Eur.\ Phys.\ J.\ C {\bf 79}, 393 (2019).


\bibitem{He:2019rva}
  J.~He and D.~Y.~Chen,
  Eur.\ Phys.\ J.\ C {\bf 79}, 887 (2019).


\bibitem{Chen:2019bip}
  H.~X.~Chen, W.~Chen, and S.~L.~Zhu,
  Phys.\ Rev.\ D {\bf 100}, 051501 (2019).

\bibitem{Azizi:2016dhy}
  K.~Azizi, Y.~Sarac, and H.~Sundu,
  Phys.\ Rev.\ D {\bf 95}, 094016 (2017).

\bibitem{Zhang:2019xtu}
  J.~R.~Zhang,
  Eur.\ Phys.\ J.\ C {\bf 79}, 1001 (2019).

\bibitem{Wang:2015qlf}
  G.~J.~Wang, L.~Ma, X.~Liu, and S.~L.~Zhu,
  Phys.\ Rev.\ D {\bf 93}, 034031 (2016).

\bibitem{Lu:2016nnt}
  Q.~F.~L{\"u} and Y.~B.~Dong,
  Phys.\ Rev.\ D {\bf 93}, 074020 (2016).


\bibitem{Xiao:2019mvs}
  C.~J.~Xiao, Y.~Huang, Y.~B.~Dong, L.~S.~Geng, and D.~Y.~Chen,
  Phys.\ Rev.\ D {\bf 100}, 014022 (2019).

\bibitem{Lin:2019qiv}
  Y.~H.~Lin and B.~S.~Zou,
  Phys.\ Rev.\ D {\bf 100}, 056005 (2019).


\bibitem{Shen:2016tzq}
  C.~W.~Shen, F.~K.~Guo, J.~J.~Xie, and B.~S.~Zou,
  Nucl.\ Phys.\ A {\bf 954}, 393 (2016).

\bibitem{Lin:2017mtz}
  Y.~H.~Lin, C.~W.~Shen, F.~K.~Guo, and B.~S.~Zou,
  Phys.\ Rev.\ D {\bf 95}, 114017 (2017).

\bibitem{Guo:2019fdo}
  F.~K.~Guo, H.~J.~Jing, U.~G.~Mei\ss{}ner, and S.~Sakai,
  Phys.\ Rev.\ D {\bf 99}, 091501 (2019).


\bibitem{Wang:2019hyc}
  Z.~G.~Wang and X.~Wang,
  Chin.\ Phys.\ C {\bf 44}, 103102 (2020).


\bibitem{Xu:2019zme}
  Y.~J.~Xu, C.~Y.~Cui, Y.~L.~Liu, and M.~Q.~Huang,
  Phys.\ Rev.\ D {\bf 102}, 034028 (2020).

\bibitem{Dong:2020nwk}
  Y.~Dong, P.~Shen, F.~Huang, and Z.~Zhang,
  Eur.\ Phys.\ J.\ C {\bf 80}, 341 (2020).

\bibitem{Wang:2019krd}
  X.~Y.~Wang, X.~R.~Chen, and J.~He,
  Phys.\ Rev.\ D {\bf 99}, 114007 (2019).


\bibitem{Wu:2019rog}
  Q.~Wu and D.~Y.~Chen,
  Phys.\ Rev.\ D {\bf 100}, 114002 (2019).

\bibitem{Wang:2019dsi}
  X.~Y.~Wang, J.~He, X.~R.~Chen, Q.~Wang, and X.~Zhu,
  Phys.\ Lett.\ B {\bf 797}, 134862 (2019).



\bibitem{Yan:2021nio}
M.~J.~Yan, F.~Z.~Peng, M.~S\'anchez S\'anchez, and M.~Pavon Valderrama,
Eur. Phys. J. C \textbf{82}, 574 (2022).

\bibitem{Nakamura:2021dix}
S.~X.~Nakamura, A.~Hosaka, and Y.~Yamaguchi,
Phys. Rev. D \textbf{104}, L091503 (2021).

\bibitem{Wang:2021crr}
J.~Z.~Wang, X.~Liu, and T.~Matsuki,
Phys. Rev. D \textbf{104}, 114020 (2021).

\bibitem{Hofmann:2005sw}
  J.~Hofmann and M.~F.~M.~Lutz,
  Nucl.\ Phys.\ A {\bf 763}, 90 (2005).

\bibitem{Wu:2010vk}
  J.~J.~Wu, R.~Molina, E.~Oset, and B.~S.~Zou,
  Phys.\ Rev.\ C {\bf 84}, 015202 (2011).

\bibitem{Wang:2019nvm}
  B.~Wang, L.~Meng, and S.~L.~Zhu,
  Phys.\ Rev.\ D {\bf 101}, 034018 (2020).

\bibitem{Anisovich:2015zqa}
  V.~V.~Anisovich, M.~A.~Matveev, J.~Nyiri, A.~V.~Sarantsev, and A.~N.~Semenova,
  Int.\ J.\ Mod.\ Phys.\ A {\bf 30}, 1550190 (2015).


\bibitem{Wang:2015wsa}
  Z.~G.~Wang,
  Eur.\ Phys.\ J.\ C {\bf 76}, 142 (2016).


\bibitem{Park:2018oib}
  W.~Park, S.~Cho, and S.~H.~Lee,
  Phys.\ Rev.\ D {\bf 99}, 094023 (2019).

\bibitem{Feijoo:2015kts}
  A.~Feijoo, V.~K.~Magas, A.~Ramos, and E.~Oset,
  Eur.\ Phys.\ J.\ C {\bf 76}, 446 (2016).

\bibitem{Lu:2016roh}
  J.~X.~Lu, E.~Wang, J.~J.~Xie, L.~S.~Geng, and E.~Oset,
  Phys.\ Rev.\ D {\bf 93}, 094009 (2016).

\bibitem{Chen:2015sxa}
  H.~X.~Chen, L.~S.~Geng, W.~H.~Liang, E.~Oset, E.~Wang, and J.~J.~Xie,
  Phys.\ Rev.\ C {\bf 93}, 065203 (2016).

\bibitem{Shen:2020gpw}
  C.~W.~Shen, H.~J.~Jing, F.~K.~Guo, and J.~J.~Wu,
  Symmetry {\bf 12}, 1611 (2020).

\bibitem{LHCb:2020jpq}
R.~Aaij \textit{et al.} (LHCb Collaboration),
Sci. Bull. \textbf{66}, 1278 (2021).

\bibitem{Collaboration:2022boa}
R.~Aaij \textit{et al.} (LHCb Collaboration),
Phys. Rev, Lett. {\bf 131},031901 (2023).

\bibitem{Chen:2020uif}
  H.~X.~Chen, W.~Chen, X.~Liu, and X.~H.~Liu,
 Eur. Phys .C 81, 409 (2021).


\bibitem{Chen:2020opr}
  H.~X.~Chen,
 Chin, Phys.C {\bf 46},093105 (2022).

\bibitem{Chen:2020kco}
  R.~Chen,
 Phys.Rey, D {\bf 103},054007 (2021).

\bibitem{Liu:2020hcv}
  M.~Z.~Liu, Y.~W.~Pan, and L.~S.~Geng,
  Phys. Rev. D {\bf 103},034003 (2021).

\bibitem{Feijoo:2022rxf}
A.~Feijoo, W.~F.~Wang, C.~W.~Xiao, J.~J.~Wu, E.~Oset, J.~Nieves, and B.~S.~Zou,
Phys. Lett. B \textbf{839}, 137760 (2023).

\bibitem{Karliner:2022erb}
M.~Karliner and J.~L.~Rosner,
Phys. Rev. D \textbf{106}, 036024 (2022).

\bibitem{Wang:2022mxy}
F.~L.~Wang and X.~Liu,
 Phys. Lett. B \textbf{835}, 137583 (2022).

\bibitem{Yan:2022wuz}
M.~J.~Yan, F.~Z.~Peng, M.~S\'anchez S\'anchez, and M.~Pavon Valderrama,
Phys. Rev. D \textbf{107}.074025 (2023).

\bibitem{Ozdem:2022kei}
U.~\"Ozdem,
Phys. Lett, B \textbf{836},137635 (2023).

\bibitem{Wang:2022tib}
F.~L.~Wang, H.~Y.~Zhou, Z.~W.~Liu, and X.~Liu,
Phys. Rev. D \textbf{106}, 054020 (2022).

\bibitem{Peng:2020hql}
F.~Z.~Peng, M.~J.~Yan, M.~S\'anchez S\'anchez, and M.~P.~Valderrama,
Eur. Phys. J. C \textbf{81}, 666 (2021).


\bibitem{Xiao:2021rgp}
C.~W.~Xiao, J.~J.~Wu, and B.~S.~Zou,
Phys. Rev. D \textbf{103}, 054016 (2021).


\bibitem{Zhu:2021lhd}
J.~T.~Zhu, L.~Q.~Song, and J.~He,
Phys. Rev. D \textbf{103}, 074007 (2021).

\bibitem{Burns:2022uha}
T.~J.~Burns and E.~S.~Swanson,
Phys. Lett. B \textbf{838}, 137715 (2023).

\bibitem{Nakamura:2022jpd}
S.~X.~Nakamura and J.~J.~Wu,
Phys. Rev D \textbf{108}, L011501 (2023).

\bibitem{Meng:2022wgl}
L.~Meng, B.~Wang. and S.~L.~Zhu,
Phys. Rev. D \textbf{107}, 014005 (2023).

\bibitem{Wu:2021caw}
Q.~Wu, D.~Y.~Chen, and R.~Ji,
Chin. Phys. Lett. \textbf{38}, 071301 (2021).

\bibitem{Cheng:1996cs}
  H.~Y.~Cheng,
  Phys.\ Rev.\ D {\bf 56}, 2799 (1997); {\bf 99}, 079901 (2019).


\bibitem{Li:2012cfa}
  H.~n.~Li, C.~D.~Lu, and F.~S.~Yu,
  Phys.\ Rev.\ D {\bf 86}, 036012 (2012).

\bibitem{Carrasco:2014poa}
  N.~Carrasco {\it et al.},
  Phys.\ Rev.\ D {\bf 91}, 054507 (2015).

\bibitem{Azevedo:2003qh}
  R.~S.~Azevedo and M.~Nielsen,
  Phys.\ Rev.\ C {\bf 69}, 035201 (2004).


\bibitem{Zou:2002yy}
  B.~S.~Zou and F.~Hussain,
  Phys.\ Rev.\ C {\bf 67}, 015204 (2003).


\bibitem{Weinberg:1965zz}
S.~Weinberg,
Phys. Rev. \textbf{137}, B672 (1965).

\bibitem{Baru:2003qq}
V.~Baru, J.~Haidenbauer, C.~Hanhart, Y.~Kalashnikova, and A.~E.~Kudryavtsev,
Phys. Lett. B \textbf{586}, 53 (2004).

\bibitem{Cheng:2004ru}
H.~Y.~Cheng, C.~K.~Chua, and A.~Soni,
Phys. Rev. D \textbf{71}, 014030 (2005).

\bibitem{Tornqvist:1993vu}
  N.~A.~Tornqvist,
  Nuovo Cimento A {\bf 107}, 2471 (1994).


\bibitem{Locher:1993cc}
  M.~P.~Locher, Y.~Lu, and B.~S.~Zou,
  Z.\ Phys.\ A {\bf 347}, 281 (1994).

\bibitem{Li:1996yn}
  X.~Q.~Li, D.~V.~Bugg, and B.~S.~Zou,
  Phys.\ Rev.\ D {\bf 55}, 1421 (1997).


 \end{thebibliography}
\end{document}